# Validation of Geant4 Electron Pair Production by Photons

Marcia Begalli, Gabriela Hoff, Maria Grazia Pia, Paolo G. Saracco

*Abstract*–**The first results of a project in progress for the validation of the simulation of electron-positron pair production are presented. They concern the pair production cross section in a low energy range close to the production threshold. The results hint to effects due to the granularity of tabulated cross sections.**

## I. INTRODUCTION

PHOTON conversion has been the object of theoretical and experimental interest for several decades; an extensive review can be found in [1].

All general purpose Monte Carlo codes, such as the EGS5 [2], EGSnrc [3], FLUKA [4][5], Geant4 [6][7], ITS [8], MCNP [9] and Penelope [10], deal with the simulation of photon conversion, as well as specialized codes such as CORSIKA [11] and AIRES [12] for cosmic ray experiments.

Geant4 describes pair production in the nuclear field and in the atomic electron field (more commonly known as triplet production); it accounts for the Landau-Pomeranchuk-Migdal (LPM) effect, which is responsible for the suppression of pair production at high energies. Geant4 provides a variety of models for this purpose; nevertheless, limited documentation is available in the literature regarding the validation of these models and their relative characteristics of physical accuracy and computational performance.

An investigation is in progress [13][14][15] to evaluate the physics models for the simulation of photon interactions used in general purpose Monte Carlo codes, as well as alternative calculations documented in the literature. For this purpose modeling features, such as total and differential cross sections, are compared with extensive collections of experimental data by means of statistical methods. The results of this process contribute to identify objectively the state of the art in this simulation domain, thus providing guidance for the simulation of experimental scenarios. In this context, this paper summarizes some preliminary results concerning the initial phase of validation of electron pair production cross sections, which is focused on the low energy range close to the production threshold.

## II. OVERVIEW OF THE VALIDATION PROCESS

The validation process adopts a similar strategy as described in [13].

The initial phase of the project is mainly concerned with the collection of a wide sample of experimental data suitable for the validation of total cross section models, the acquisition, digitization and reformatting of relevant data libraries, and the refactoring [16] of Geant4 pair production models into an agile software design [13].

Experimental data are collected from the literature. Those only available in figures are digitized by means of Plot Digitizer [17]; the digitization error is estimated and taken into account. Systematic effects in the data are identified whenever possible; outliers are identified and excluded from the statistical analysis.

In the initial phase of the validation process a limited set of total pair production cross section modeling methods has been investigated. It encompasses:

- EPDL [18] tabulations, used by Geant4 so-called "Livermore" models,
- the parameterization implemented in Geant4 *G4BetheHeitlerModel* class,
- the tabulated cross sections used in Penelope 2008 and 2011 versions: a re-engineered implementation of Penelope 2008 pair production simulation is available in Geant4 *G4PenelopeGammaConversionModel* class,
- Storm and Israel's [19] tabulations: they are available as an optional cross section source in EGSnrc,
- XCOM [20] tabulations, on which Penelope cross sections are based.

With the exception of Storm and Israel's tabulations, all the above mentioned cross sections are based on the calculations by Hubbell, Gimm and Øverbø [21]; differences may arise in the cross section calculated in the course of particle transport as effects of different energy grids at which cross sections have been tabulated or, as in the case of Geant4 *G4BetheHeitlerModel* class, of parameterization of tabulated values.

These total cross sections are implemented in two policy classes [22], responsible for the interpolation of tabulated cross sections and for the calculation of parameterized cross

Manuscript received November 15, 2013.
This work has been partly funded by INFN grant, Italy.
M. Begalli is with State University of Rio de Janeiro, Brazil (e-mail: begalli@cern.ch).
This work has been partly funded by INFN grant, Italy.
M. Begalli is with State University of Rio de Janeiro, Brazil (e-mail: begalli@cern.ch).
G. Hoff is with Pontificia Universidade Catolica do Rio Grande do Sul, Brazil (e-mail: Gabriela.Hoff@ge.infn.it)
G. Hoff is with Pontificia Universidade Catolica do Rio Grande do Sul, Brazil (e-mail: Gabriela.Hoff@ge.infn.it)
M. G. Pia and Paolo Saracco are with INFN Sezione di Genova, Via Dodecaneso 33, I-16146 Genova, Italy (phone: +39 010 3536328, fax: +39010 313358, MariaGrazia.Pia@ge.infn.it, Paolo.Saracco@ge.infn.it).

sections according to the algorithm derived from Geant4 *G4BetheHeitlerModel* class, respectively.

For the validation of simulation methods, cross sections are calculated in the same configuration (target atom and incident photon energy) as in the experimental data sample. They are compared with experimental data by means of the $\chi^2$ test [23]. The null hypothesis in the test consists of the equivalence of experimental and calculated distributions; a significance level $\alpha = 0.01$ is defined for the rejection of the null hypothesis.

### III. PRELIMINARY RESULTS

The validation analysis in the initial phase of the project has focused on energies close to the production threshold. It is based on a data sample of approximately one hundred measurements at energies between 1.06 and 2.62 MeV. Some examples of experimental and calculated cross sections are shown in Fig. 1 and 2.

The distribution of the difference between calculated and experimental cross sections, expressed in terms of number of experimental standard deviations, is plotted in Fig. 3.

Table I reports the p-values resulting from the $\chi^2$ test. In the table G4std identifies the parameterized cross section refactored from the Geant4 G4BetheHeitlerModel class. The null hypothesis of compatibility of the calculated cross sections is rejected in all test cases, with the exception of the cross sections based on EPDL above 1.12 MeV.

TABLE I. P-VALUES OF THE $\chi^2$ TEST

| Energy (MeV) | G4std | EPDL | XCOM | Penelope 2008 | Penelope 2011 | Storm Israel |
|---|---|---|---|---|---|---|
| <1.12 | <0.001 | <0.001 | <0.001 | <0.001 | <0.001 | <0.001 |
| 1.12-2.62 | <0.001 | 0.982 | <0.001 | <0.001 | <0.001 | <0.001 |

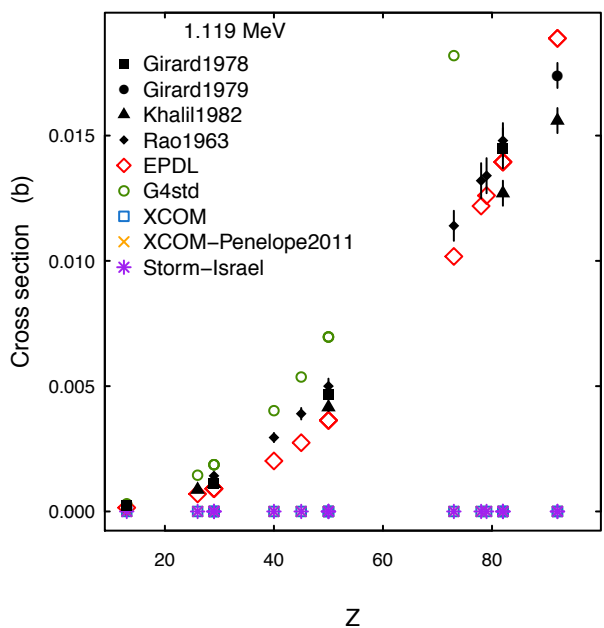

Fig. 1. Experimental (black filled symbols) and calculated (open symbols) pair production cross sections at 1.119 MeV photon energy, as a function of atomic number. G4std identifies the parameterized cross section refactored from the Geant4 *G4BetheHeitlerModel* class.

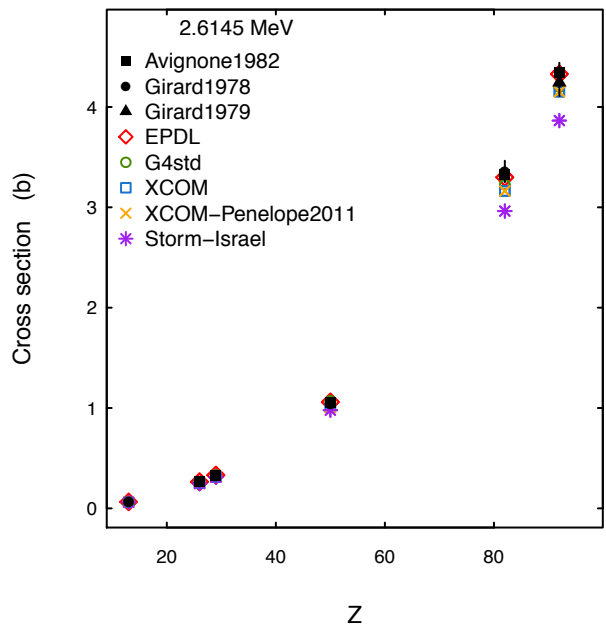

Fig. 2. Experimental (black filled symbols) and calculated (open symbols) pair production cross sections at 1.119 MeV photon energy, as a function of atomic number. G4std identifies the parameterized cross section refactored from the Geant4 *G4BetheHeitlerModel* class.

### IV. CONCLUSION AND OUTLOOK

An investigation is in progress to evaluate quantitatively models for the simulation of electron-positron pair production against a large collection of experimental data. It is part of a wider project for the validation of photon interactions in Monte Carlo particle transport codes.

The initial phase of the project is focused on the validation of total pair production cross sections at energies close to the threshold. Preliminary results indicate that cross sections based on EPDL tabulations are compatible with experimental data in the energy range between 1.12 and 2.62 MeV with 0.01 significance. Cross sections based on XCOM tabulations and on a parameterization implemented in Geant4 "standard" electromagnetic package do not appear to be compatible with experiment in the same energy range. Since all these cross section calculations are based on the same source [21], the observed discrepancy in their compatibility with experimental data could be attributed to effects related to the granularity of the tabulation grid or to the precision of the parameterization.

More extensive results of the validation tests currently in progress will be reported in a forthcoming publication.